# Atomic-scale spin-polarization maps using functionalized superconducting probes

Lucas Schneider[1], Philip Beck[1], Jens Wiebe[1,*] and Roland Wiesendanger[1]

[1]Department of Physics, University of Hamburg, D-20355 Hamburg, Germany.

*Correspondence to: jwiebe@physnet.uni-hamburg.de

**A scanning tunneling microscope (STM) with a magnetic tip that has a sufficiently strong spin-polarization can be used to map the sample's spin structure down to the atomic scale but usually lacks the possibility to absolutely determine the value of the sample's spin-polarization. Magnetic impurities in superconducting materials give rise to pairs of perfectly, i.e. 100% spin-polarized sub-gap resonances. In this work, we functionalize the apex of a superconducting Nb STM-tip with such impurity states by attaching Fe atoms to probe the spin-polarization of atom-manipulated Mn nanomagnets on a Nb(110) surface. By comparison with spin-polarized STM measurements of the same nanomagnets using Cr bulk tips we demonstrate an extraordinary spin-sensitivity and the possibility to measure the sample's spin-polarization values quantitatively with our new functionalized probes.**

**One Sentence Summary:** Functionalized superconducting STM tips can be used to map the spin polarization distribution of nanomagnets quantitatively and with atomic spatial resolution.

Spin-polarized scanning tunneling microscopy (SP-STM) and spectroscopy (SP-STS) exploit the tunneling magnetoresistance between two spin-polarized electrodes, the sample and the STM tip, separated by a vacuum barrier (*1*) as sketched in Fig. 1A. A bias voltage $V$ is applied, shifting the Fermi energies $E_\mathrm{F}$ of sample and tip by an amount $eV$. This results in a net current, which is large in the case of parallel alignment (↑↑) of the dominating spin species in the density of states (DOS) of the tip and in the local density of states (LDOS) of the sample, evaluated at the tip apex position $\vec{r}_t$ in the energy window of tunneling, but small for antiparallel alignment (↑↓), realized, e.g., by using an external magnetic field $B_z$. The resulting asymmetry $A_\mathrm{SP}(V,\vec{r}_t)$ in the measured spin-resolved differential tunneling conductance $\mathrm{d}I/\mathrm{d}V\,(V)$ is approximately given by the product of the spin-polarizations of tip $\mathcal{P}_t(E_\mathrm{F})$ and sample $\mathcal{P}_s(E_\mathrm{F}+eV,\vec{r}_t)$ (*2*):

$$A_\mathrm{SP}(V,\vec{r}_t) = \frac{\mathrm{d}I/\mathrm{d}V(V,\vec{r}_t)|_{\uparrow\uparrow} - \mathrm{d}I/\mathrm{d}V(V,\vec{r}_t)|_{\uparrow\downarrow}}{\mathrm{d}I/\mathrm{d}V(V,\vec{r}_t)|_{\uparrow\uparrow} + \mathrm{d}I/\mathrm{d}V(V,\vec{r}_t)|_{\uparrow\downarrow}} \approx \mathcal{P}_t(E_\mathrm{F}) \cdot \mathcal{P}_s(E_\mathrm{F}+eV,\vec{r}_t) \qquad (1)$$

This emphasizes the crucial role of a large tip spin-polarization. Along these lines a large number of studies using different tip materials have been performed (*1*). These include half-metallic ferromagnets (*3*), bulk 3*d* transition metals (*4*,*5*), thin-films of 3*d* transition metals (*6*), or soft magnetic tips periodically magnetized in opposite directions along the tip axis by a small coil (*7*). For instance, Cr tips feature a low stray-field, but the values for their spin-polarization are only as large as 10 - 20%, depending on the specific tip apex (*1*). Additionally, since $\mathcal{P}_t(E_\mathrm{F})$ is usually a priori unknown for most of the common materials, conventional SP-STS cannot be used straight-forwardly to measure absolute values of $\mathcal{P}_s(E_\mathrm{F}+eV,\vec{r}_t)$. Tip materials with a well-known spin-polarization $\mathcal{P}_t(E_\mathrm{F})$ can be either hard to implement experimentally (such as optically pumped GaAs (*8*,*9*), requiring optical access to the tunneling junction) or require large magnetic fields on the order of a few Tesla, thereby perturbing the sample's magnetic state considerably (*10*). Finally, it is not straight-forward to disentangle the magnetoresistive change in tunneling conductance from locally varying electronic contrast of the sample.

Yu-Shiba-Rusinov (YSR) states induced by magnetic impurities in superconductors (*11*–*13*) have gained renewed interest in the past years, mainly with the focus to realize topological superconductors and Majorana bound states in one- or two-dimensional arrays of such impurities (*14*–*19*). Yet another interesting feature of these bound states is their perfect, i.e. 100 % spin-polarization (*20*–*22*), naturally raising the question whether it is possible to use them as a probe for the spin-polarization of an unknown sample under investigation.

Along these lines, our idea of a novel "YSR-SP-STM" method is sketched in Fig. 1B: A pair of entirely spin-polarized YSR subgap resonances can appear in the density of states of a superconducting tip, which is experimentally realized by picking up magnetic atoms to the tip apex, and magnetizing these atoms with an external magnetic field $B_z$ (*23*). The energetic position of both peaks is determined by the exchange interaction strength of the impurity spin with the Cooper pairs in the superconductor. Unlike on an ordered surface, the adsorption positions of adatoms on an STM tip can be manifold. It is therefore possible to realize almost arbitrary energetic positions of the tip YSR states. They always appear in pairs at $E_\text{F} \pm \varepsilon$ with a particle-like state (labeled $e$ and ↑ in the following) and a hole-like state (labeled $h$ and ↓). With increasing exchange coupling of the impurity to the host superconductor, the YSR states shift in energy, starting from the superconducting gap edge $\Delta_\text{tip}$. They eventually cross at $E_\text{F}$ and shift towards $-\Delta_\text{tip}$ again. At the crossing point, they undergo a quantum phase transition (QPT) (*14*, *21*, *24*).

On the other side of the junction, the partially spin-polarized LDOS of the sample is probed. For the case $e \cdot V = E_{\text{YSR},\uparrow}$ (Fig. 1B, upper panel), tunneling from the ↑ YSR state to the sample is allowed and results in a large tunneling conductance if this state and the spins dominating the sample's LDOS are aligned (↑↑). With the reversed voltage $e \cdot V = E_{\text{YSR},\downarrow}$, tunneling in this setup from the sample to the ↓ YSR state in the tip is less probable since there are fewer spin-down electrons available in the sample's LDOS. In consequence, the asymmetry $A_\text{YSR}(\vec{r}_t)$ in the intensities $a_e$ and $a_h$ of the two YSR peaks observed in $\mathrm{d}I/\mathrm{d}V$ measurements, will change according to their spin's relative alignment to the local dominating sample spin-direction according to

$$A_\text{YSR}(\vec{r}_t) = \frac{a_e(\vec{r}_t) - a_h(\vec{r}_t)}{a_e(\vec{r}_t) + a_h(\vec{r}_t)} \approx \mathcal{P}_s(E_\text{F}, \vec{r}_t) \tag{2}$$

In the last equation, we have hypothesized a perfectly spin-polarized pair of YSR states and assumed that the LDOS is largely energy independent around $E_\text{F}$, which is the case for many sample systems of interest. The technique, thus, features several advances compared to traditional SP-STM: most notably, the perfect spin-polarization of the probe gives maximized spin contrast and can allow for a quantitative measurement of spin-polarization of the vacuum LDOS above the sample. It also facilitates the distinction of spin-contrast and electronic contrast, as the asymmetry of peaks $A_\text{YSR}$ on the µeV-scale is unlikely to be strongly affected by a local change in the non-magnetic LDOS. Furthermore, it enables the use of superconducting tips providing excellent energy resolution below the Fermi-Dirac limit for spin-polarized measurements. In the following, we experimentally demonstrate that we can indeed realize this technique with a pair of highly spin-polarized YSR states using a Nb-coated W-tip with Fe atoms attached to the apex as a YSR probe and a clean Nb(110) surface with Mn nanomagnets as a sample.

All experiments were performed under ultra-high vacuum conditions ($p < 2 \cdot 10^{-10}$ mbar) in a home-built STM setup at a temperature of $T$ = 300 mK (*25*). We study a clean Nb(110) surface (*26*) with Mn and Fe adatoms to characterize the magnetic imaging of our STM tips. Additional information on the preparation of the sample can be found in (*27*). The Mn and Fe adatoms adsorb on the four-fold coordinated hollow sites of the (110) lattice of surface Nb atoms, as sketched in Fig. 2A. Using STM-tip induced single atom manipulation at typical tunneling resistances of $R$ = 60 kΩ, the Mn atoms can be moved to the desired positions on the Nb(110) surface. In this way, artificial chains with arbitrary inter-atomic distances and in different crystallographic directions can be constructed. The magnetic ordering in such chains is expected to be strongly affected by these geometric properties, due to the oscillatory behavior of the mediating Ruderman-Kittel-Kasuya-Yosida (RKKY) interaction (*28-30*). Using the lattice constant of the (110) surface $a$, we introduce the following nomenclature for referencing to the distance of the nearest neighboring atoms in the three different investigated chains and their orientations (Fig. 2A): $1a - [001]$, $2a - [001]$ and $\frac{\sqrt{3}a}{2} - [1\bar{1}1]$. As we will prove in the following experimentally via SP-STS using bulk Cr tips (see (*27*) for tip preparation procedures), $2a - [001]$ and $\frac{\sqrt{3}a}{2} - [1\bar{1}1]$ chains exhibit an antiferromagnetic spin structure, whereas the $1a - [001]$ chains are ferromagnetic.

As a first example, a $2a - [001]$ Mn$_5$ chain was constructed (Fig. 2B). Such arrays of weakly RKKY-coupled atoms on metallic surfaces are known to have a very short lifetime of degenerate spin states and we do not expect any magnetic contrast in zero-field (*29*). Therefore, in order to stabilize the spin structure of the chains, an external magnetic field of -0.5 T is applied perpendicular to the sample surface in ↓ direction. Assuming an antiferromagnetic chain consisting of an odd number of atoms, the ↑↓↑↓↑ spin configuration is expected to be favored over the ↓↑↓↑↓ configuration by Zeeman energy. Indeed, we find an alternating contrast along the chain when mapping the differential tunneling conductance d$I$/d$V$ (Fig. 2B). Reversing the magnetic field polarity also inverts the contrast as expected if we have a stable magnetization of the tip while the chain magnetization aligns paramagnetically with the field. This behavior is thus consistent with the assumption of a chain of antiferromagnetically RKKY-coupled atoms (*29*). The total change in magnetic contrast can be seen in the asymmetry map $A_\text{SP}(V, \vec{r}_t)$ on the right side of Fig. 2B. Neighboring atoms clearly exhibit alternating sign in the spin-asymmetry, proving antiferromagnetic coupling between the neighboring atoms for this particular chain type. In the same fashion, $\frac{\sqrt{3}a}{2} - [1\bar{1}1]$ Mn$_{13}$ chains (Fig. 2C) and $1a - [001]$ Mn$_{15}$ chains (Fig. 2D) have been studied. For the former, we again find an alternating contrast between adjacent atoms, which is inverse when reversing the external magnetic field, clearly indicating a dominant antiferromagnetic nearest neighbor coupling in the chain. The latter exhibits no clear contrast along the chain (Fig. 2D). However, the d$I$/d$V$ signal changes from bright to dark in $B_z = \pm 0.5$ T. This is apparently visible in the according asymmetry map, revealing a ferromagnetic ordering along this chain type.

With this knowledge about the spin structures of the various chains, we can probe similar structures with a tip hosting YSR states in a next step. Superconducting tips were obtained by indenting an electrochemically etched W-tip into the Nb(110) surface for several nanometers. The resulting superconducting cluster of the tip exhibits a much higher critical field $B_c$ than the sample due to its finite size and it is only weakly affected by the applied magnetic fields of $B_z = 0.5$ T (see Fig. 3B, gray, and Supplementary Fig. S3) whereas the Nb(110) sample is in the normal conducting state ($B_{c,2,Nb} = 0.4$ T (*31*)). It is even preferable to keep the sample in the normal conducting state for this experiment since vortices or YSR states in the sample would strongly hamper the interpretation of d$I$/d$V$ spectra obtained with a YSR tip. Subsequently, magnetic atoms were picked up from the sample surface by lowering the tip close to the atom, applying a voltage pulse of 2 V and retracting the tip back to normal tunneling conditions. One example can be seen in Fig. 3A, showing a series of images where two Fe atoms are transferred to the tip. The atoms are likely to be located on the tip apex as the imaging quality of the tip is improved, comparing the observed shape of the Mn atom between the right and the left image. As a result, the tip DOS now features two additional in-gap states (Fig. 3B, red). Assuming that the Nb substrate in the normal state features a nearly flat LDOS within a few meV around the Fermi level, we can conclude that we measure only the tip DOS in the d$I$/d$V$ spectroscopy to a good approximation. These states can thus be interpreted as YSR states of the Fe$_2$ impurity locally perturbing the superconducting tip material. Note, that the particle- and hole-like peaks in these spectra measured on the nonmagnetic substrate already have an intrinsic asymmetry in the intensities, which we name $a_e(\vec{r}_t = \text{sub})$ and $a_h(\vec{r}_t = \text{sub})$, respectively. This is a well-known effect of an additional non-magnetic scattering term at the impurity (*20, 21, 32*). The intrinsic asymmetry is quantitatively evaluated by a fit of two Lorentzian distributions to the sub-gap energy values in the d$I$/d$V$ spectra (see Supplementary Fig. S2). The resulting peak heights have a ratio of $a_e(\vec{r}_t = \text{sub}) / a_h(\vec{r}_t = \text{sub}) = 1.82 \pm 0.04$.

With this tip, we were able to assemble various magnetic Mn chains, starting with a structurally identical $2a - [001]$ chain as it has been presented in Fig. 2B. A d$I$/d$V$ map at $B_z = 0.5$ T (Fig. 3C) reproduces the clear alternating AFM contrast between neighboring chain atoms, demonstrating a strong out-of-plane magnetic contrast of the tip. d$I$/d$V$ spectra of the tip YSR states obtained with this tip stabilized above the five chain atoms (Fig. 3D) reveal a strong change in asymmetry of the peak heights when tunneling between the tip's YSR states and the Mn atoms along the chain magnetized in opposing directions. This asymmetry can now directly be attributed to spin-polarized tunneling. In the following we define the YSR asymmetry $A_{\text{YSR,norm}}(\vec{r}_t)$, which has been normalized in order to account for the intrinsic YSR asymmetry, as:

$$A_{\text{YSR,norm}}(\vec{r}_t) = \frac{a_e(\vec{r}_t)/a_e(\vec{r}_t = \text{sub}) - a_h(\vec{r}_t)/a_h(\vec{r}_t = \text{sub})}{a_e(\vec{r}_t)/a_e(\vec{r}_t = \text{sub}) + a_h(\vec{r}_t)/a_h(\vec{r}_t = \text{sub})} \qquad (3)$$

From the previous study with the Cr tip we know that the spin structure of the chain is anti-aligned with the external field and reads ↓↑↓↑↓ for positive $B_z$. The tip apex consists of only two magnetic atoms on a metallic tip material whose spin will probably fully anti-align with a field of strength $B_z = \pm 0.5$ T as well (↓) (*33*). A plot of $A_{\text{YSR,norm}}$ against the spatial position on the chain is shown in Fig. 3E. For parallel alignment of tip and sample magnetization, which is the case for atoms 1, 3 and 5, we find a normalized YSR peak asymmetry of $A_{\text{YSR,norm}}(1,3,5) = (15.2 \pm 0.5)\%$. For antiparallel alignment on the atoms 2 and 4, the asymmetry correspondingly equals $A_{\text{YSR,norm}}(2,4) = (-16.4 \pm 0.7)\%$. Obviously, $A_{\text{YSR,norm}}$ resembles the asymmetry map obtained with conventional SP-STM (Fig. 2B) which is plotted in the bottom panel of Fig. 3E for direct comparison. However, we find several striking differences. (i) When comparing the absolute values, we obtain a signal being larger by one order of magnitude using the YSR-SP-STS method with respect to conventional SP-STS. (ii) For the former, we can distinguish between magnetic- and non-magnetic contrast without reversing the magnetic field while this is not possible for conventional SP-STS. What remains to be proven is the atomic spatial resolution of YSR-SP-STS. The inter-atomic distance in the $2a - [001]$ Mn$_5$ chain studied in Fig. 3 equals 660 pm, which is well above typical lattice constants of investigated single crystals. Therefore, we again built the densely packed $\frac{\sqrt{3}a}{2} -$

[1$\bar{1}$1] Mn$_{15}$ (Fig. 4A, 285 pm inter-atomic distance) and $1a - $ [001] Mn$_{15}$ chains (Fig. 4B, 330 pm inter-atomic distance) with the YSR tip to compare the results with those obtained with the bulk Cr tip from Fig. 2C and D, respectively. Evaluating $A_{\text{YSR,norm}}$ for the two cases we find an alternating signal on the $\frac{\sqrt{3}a}{2} - $ [1$\bar{1}$1] Mn$_{15}$ chain (Fig. 4A), indicating antiferromagnetic ordering. Moreover, we find a homogeneous signal strength on the $1a - $ [001] Mn$_{15}$ chain, indicating ferromagnetically aligned spins (Fig. 4B). Both results reproduce the magnetic structures found in Fig. 2C and D, but yield much larger signal magnitudes, thereby demonstrating that significant spin contrast can be obtained by the YSR-SP-STS technique even for densely packed structures on the atomic scale.

We finally discuss the proposed possibility to measure the absolute values of the sample's spin-polarization using YSR-SP-STS data (Eq. 2). Assuming that the magnetic field of strength $B_z = \pm 0.5$ T was indeed able to fully saturate the Fe$_2$ cluster on the tip, the measured spin-polarization on the $2a - $ [001] Mn$_5$ chain would equal $\mathcal{P}_s(E_F, \vec{r}_t) = 15.8\%$. Considering Eq. 1 and the asymmetry measured by conventional SP-STS on the same chain of $A_{\text{SP}}(V, \vec{r}_t) = 1.7\%$, we obtain a spin-polarization of $\mathcal{P}_t(E_F) = 9.3\%$ for the Cr tip used in Fig.2B/C which is a reasonable value for bulk Cr tips (*1*). Similar ratios are found for the examples in Fig. 4. The only reason for a non-perfect spin-polarization of the tip states in YSR-SP-STS would be the existence of multiple YSR states which cannot be distinguished due to our energy resolution (*25, 34*). If the latter states were on different sides of the QPT, they would have opposite spin-polarizations and the net spin-polarization in a YSR-SP-STS measurement would be reduced. As we cannot exclude that our tip features multiple YSR states, further progress can be made in the design of tips featuring only a single pair of YSR states (*35*) for obtaining even higher spin-polarization, e.g. using other impurities than transition metal atoms which typically induce a multiplet of non-degenerate peaks due to their *d*-orbital magnetism (*24, 36, 37*).

In summary, our results clearly demonstrate the advantages of using superconducting STM tips functionalized by magnetic impurity bound states to measure atomic-scale spin textures. Due to the inherent perfect spin polarization of YSR states, we can obtain spin-contrast with one order of magnitude larger signal intensities compared to conventional Cr bulk tips and determine absolute values of the sample's spin-polarization.

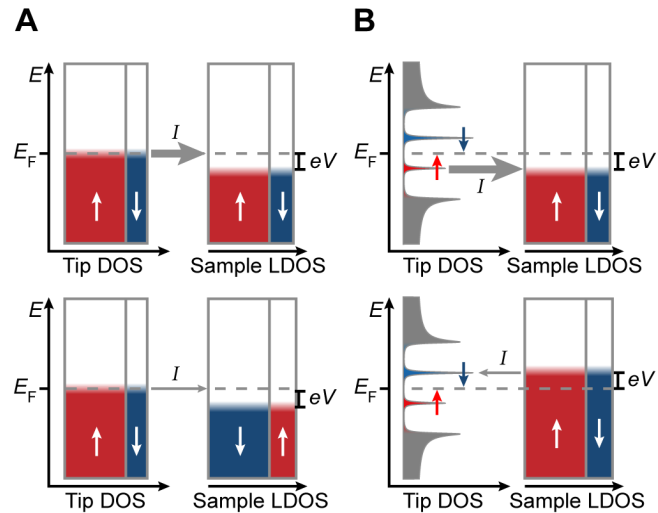

**Fig. 1 | Concept of SP-STM and YSR-SP-STM. (A)** Schematic drawing of the concept of a conventional SP-STM experiment. The spin-resolved tip DOS and sample LDOS are assumed to be energy independent for simplicity. The thickness of the vertical arrows indicate the strength of the main contribution to the tunnel current. Between the upper and the lower panel the sample magnetization has been reversed, which enables a measurement of the tunneling magnetoresistance effect. **(B)** Schematic drawing of the concept of a YSR-SP-STM experiment. A single YSR state in the gap of a superconducting tip which is magnetized, e.g. by an external magnetic field, is assumed for simplicity. Between the upper and the lower panel the bias polarity has been reversed, which enables the measurement of the absolute value of the sample's spin-polarization. Note, that there is an intrinsic asymmetry in the intensity of the ↑ YSR state with respect to the ↓ YSR state which is unrelated to magnetism.

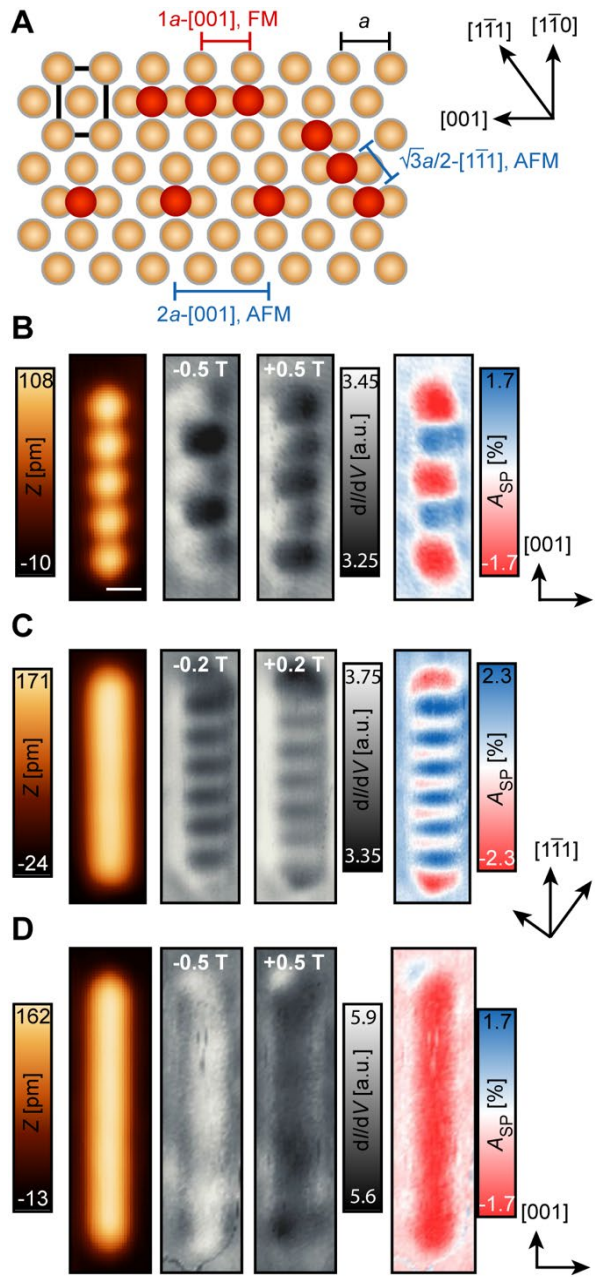

**Fig. 2 | Conventional SP-STM on chains along different crystallographic directions using a Cr bulk tip. (A)** Sketch of the uppermost Nb(110) layer (brown) with Mn adatoms (red) on hollow sites. Using STM-tip based atom-manipulation, the adatoms can be arranged with different options for inter-atomic spacings and crystallographic directions, which strongly affects their magnetic coupling (FM: ferromagnetic, AFM: antiferromagnetic). **(B)** Constant-current STM image of a $2a - [001]$ $Mn_5$ chain, the corresponding d$I$/d$V$ maps in opposite external magnetic fields $B_z$ and the calculated $A_{SP}$. **(C)** Constant-current STM image of a $\frac{\sqrt{3}a}{2} - [1\bar{1}1]$ $Mn_{13}$ chain, the corresponding d$I$/d$V$ maps in opposite external magnetic fields $B_z$ and the calculated $A_{SP}$. **(D)** Constant-current STM image of a $1a - [001]$ $Mn_{15}$ chain, the corresponding d$I$/d$V$ maps in opposite external magnetic fields $B_z$ and the calculated $A_{SP}$. The white bar in (B) corresponds to 500 pm, all images share the same scale. Parameters: ($V = -5$ mV, $I = 1$ nA, $V_{mod} = 2$ mV) for (B)/(C) and ($V = -6$ mV, $I = 1$ nA, $V_{mod} = 2$ mV) for (D).

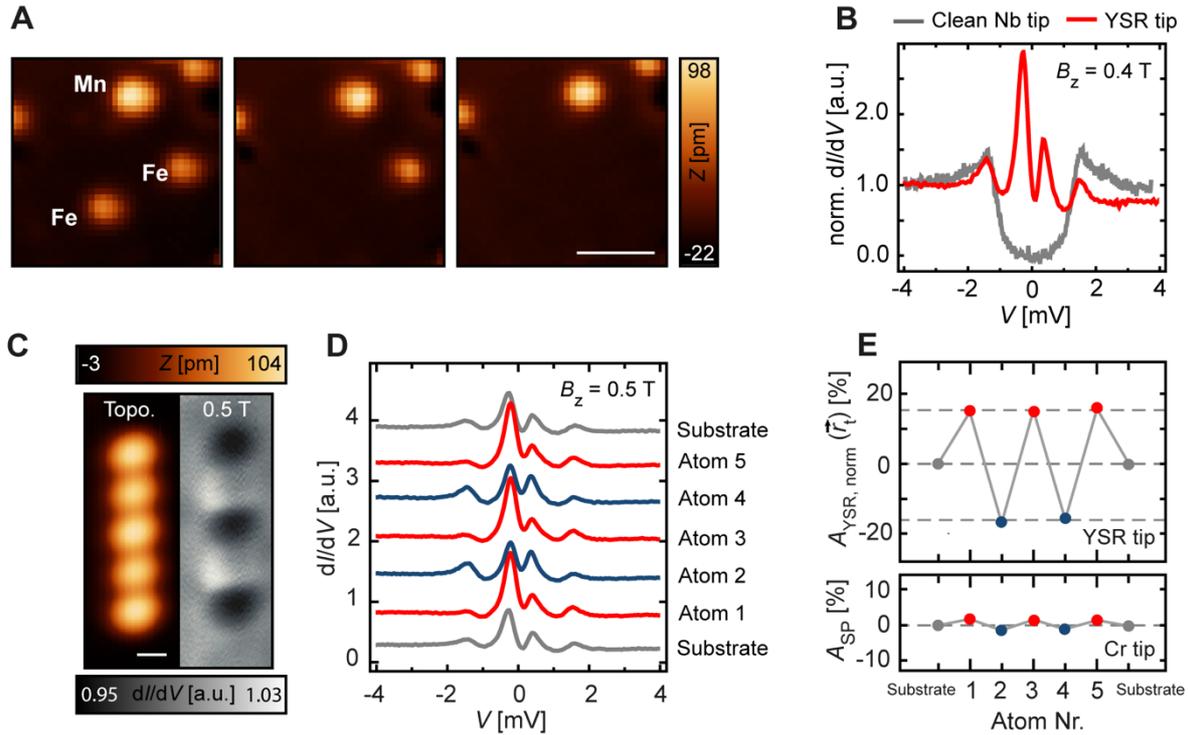

**Fig. 3 | Picking up two Fe atoms onto a Nb tip and demonstrating out-of-plane magnetic contrast. (A)** Constant-current STM images before (left two images) and after picking up the two indicated Fe atoms (right image) onto the tip ($V$ = 6 mV, $I$ = 2 nA). The white bar corresponds to 2 nm. **(B)** d$I$/d$V$ spectra taken on the normal conducting Nb(110) substrate in an external field of $B_z$ = 0.4T with (red) and without (gray) Fe atoms attached to the tip ($V_{stab}$ = −10 mV, $I_{stab}$ = 2 nA, $V_{mod}$ = 100 µV for the YSR tip and $V_{stab}$ = −6 mV, $I_{stab}$ = 1 nA, $V_{mod}$ = 20 µV for the clean Nb tip). **(C)** Constant-current STM topography (left) and the according d$I$/d$V$ map (right) of an antiferromagnetic $2a - [001]$ Mn$_5$ chain measured at $B_z$ = 0.5 T and stabilized outside of the superconducting gap ($V_{stab}$ = −10 mV, $I_{stab}$ = 2 nA, $V_{mod}$ = 3 mV). The white bar corresponds to 0.5 nm. **(D)** d$I$/d$V$ spectra taken on the substrate and on the five atoms of the chain ($V_{stab}$ = −10 mV, $I_{stab}$ = 2 nA, $V_{mod}$ = 100 µV) with the YSR tip in an external field of $B_z$ = 0.5 T. The spectra are vertically offset for clarity. **(E)** Asymmetry $A_{YSR,norm}(\vec{r}_t)$ of the YSR peaks in (D) calculated from Eq. 3 (upper panel) compared to the spin-asymmetry $A_{SP}(\vec{r}_t)$ obtained with the Cr tip (lower panel) using the data from Fig. 2B and averaging across the atoms and a substrate position. The off-zero dashed gray lines mark the average YSR Asymmetry measured on ↑ and ↓ atoms.

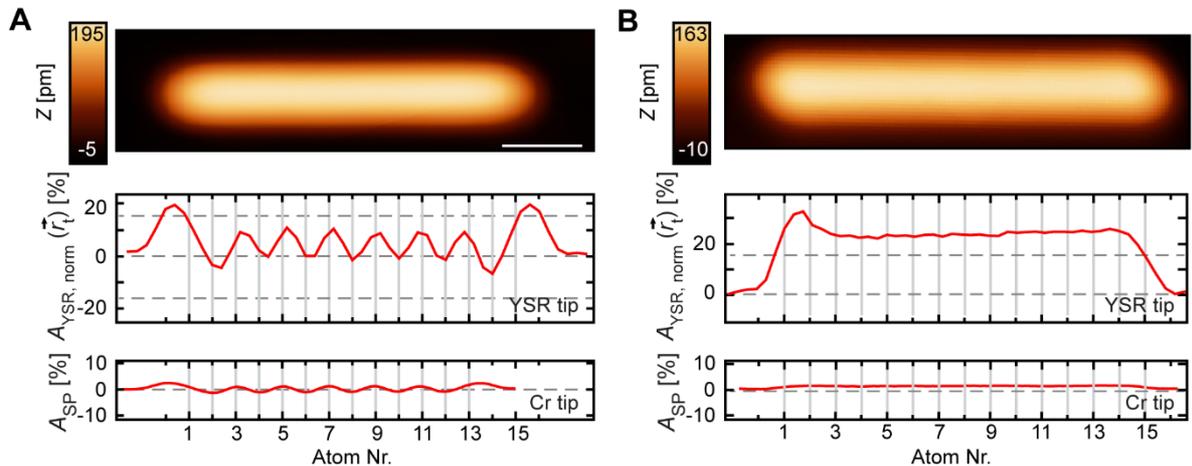

**Fig. 4 | Magnetic YSR contrast on densely-packed adatom structures. (A)** Constant-current STM image of a $\frac{\sqrt{3}a}{2} - [1\bar{1}1]$ Mn$_{15}$ chain and the YSR asymmetry measured along the chain. The lower panel shows the contrast obtained with a Cr tip on

the similar $Mn_{13}$ chain of Fig. 2C. **(B)** Constant-current STM image of a $1a - [001]$ $Mn_{15}$ chain and the YSR asymmetry measured along the chain. The lower panel shows the contrast obtained with a Cr tip on the same $Mn_{15}$ chain in Fig. 2D. The non-zero dashed gray lines mark the average YSR Asymmetry measured on ↑ and ↓ atoms from Fig. 3E. The white bar in (A) corresponds to 1 nm. All data was measured with $V_{stab}$ = −10 mV, $I_{stab}$ = 2 nA, $V_{mod}$ = 100 μV, $B_z$ = 0.5 T.

## Acknowledgements
L.S., R.W., and J.W. gratefully acknowledge funding by the Cluster of Excellence 'Advanced Imaging of Matter' (EXC 2056 - project ID 390715994) of the Deutsche Forschungsgemeinschaft (DFG). P.B., R.W., and J.W. acknowledge support by the SFB 925 'Light induced dynamics and control of correlated quantum systems' of the DFG. R.W. gratefully acknowledges financial support from the European Union via the ERC Advanced Grant ADMIRE (project No. 786020).

## Data availability
The authors declare that the data supporting the findings of this study are available within the paper and its supplementary information files.

## Competing interests
The authors declare no competing interests.

## Author contributions
L.S., J.W. and R.W. conceived the experiments. L.S. and P.B. performed the measurements. L.S. and P.B. analyzed the experimental data. L.S. and J.W. prepared the figures and wrote the manuscript. All authors contributed to the discussions and to correcting the manuscript.

## Supplementary Information
Materials and Methods
Supplementary Text
Fig. S1 – S3